\documentstyle[12pt,emulateapj,psfig]{article}

\newcommand{\gtilde}
{~ \raisebox{-1ex}{$\stackrel{\textstyle >}{\sim}$} ~}
\newcommand{\ltilde}
{~ \raisebox{-1ex}{$\stackrel{\textstyle <}{\sim}$} ~}

\begin{document}
\footnotesize

\submitted{To Appear in the Astrophysical Journal Letters}

\title{An Interpretation of the Evidence for TeV Emission from \\
Gamma-Ray Burst 970417a}

\author{Tomonori Totani}
\affil{Theory Division, National Astronomical Observatory \\
Mitaka, Tokyo 181-8588, Japan \\
e-mail: totani@th.nao.ac.jp
}

\begin{abstract}
The Milagrito collaboration recently reported evidence for 
emission of very high energy gamma-rays in the TeV range from
one of the BATSE GRBs, GRB 970417a. Here I discuss
possible interpretations of this result. Taking into account
the intergalactic absorption of TeV gamma-rays by the cosmic
infrared background, I found that the detection rate (one per 54
GRBs observed by the Milagrito) and energy fluence can be
consistently explained with the redshift of this GRB at $z \sim 0.7$
and the isotropic total energy in the TeV range,
$E_{\rm TeV, iso} \gtilde 10^{54}$ erg. This energy scale
is not unreasonably large, but interestingly similar to
the maximum total GRB energy observed to date, in the sub-MeV range
for GRB 990123.
On the other hand, the energy emitted in the ordinary sub-MeV range
becomes $E_{\rm MeV, iso} \sim 10^{51}$ erg for the GRB 970417a,
which is much smaller than the total energy in the TeV range
by a factor of about $10^3$. I show that 
the proton-synchrotron model of GRBs provides a possible 
explanation for these observational
results. I also discuss some observational signatures expected in
the future experiments from this model.
\end{abstract}


\section{Introduction}
Gamma-ray bursts (GRBs) have been the most mysterious astronomical
phenomenon in the universe for about 30 years after the discovery
(Klebesadel, Strong, \& Olson 1973). 
One of the reasons why the GRB phenomenon is difficult to
understand is that GRBs had been observed only in the soft gamma-ray band
for a long time. Detection of GRBs
in other wavelengths is quite valuable for the progress of the GRB study,
as proved by the dramatic progress in recent years following the
discovery of afterglows in longer wavelengths of X, optical and radio bands
(see, e.g., Piran 1999 for a recent review).
Detection of gamma-rays harder than the ordinary sub-MeV band 
is also important information for better understanding of GRBs.
It has been confirmed that emission from GRBs extends up to $\sim$ 10
GeV, as seen in the famous long-duration GeV emission from GRB 940217
(Hurley et al. 1994). There have been some suggestive results for the emission
beyond TeV range (Amenomori et al. 1996; Padilla et al. 1998), 
although these results were not claimed as firm detections of GRBs.

Recently the Milagro group reported evidence for TeV emission from
one (GRB 970417a) of the 54 BATSE GRBs 
in the field of view of their detector, Milagrito (proto-type of
Milagro) (Atkins et al. 2000).
An excess of gamma-rays above background is clearly seen during the
duration of this burst in the
BATSE error circle, and the chance probability 
of such an event after examining 54 GRBs is estimated as $1.5 \times 
10^{-3}$, giving stronger evidence for TeV emission 
compared with the earlier observations in this energy band.
If this signal is truly from the GRB, the TeV fluence must be at least
10 times greater than the sub-MeV fluence of this GRB without
taking into account the intergalactic absorption of TeV gamma-rays. The impact
on the GRB energetics would be quite strong. 

In this letter I discuss theoretical implications of this interesting
event, assuming that the signal observed by the Milagrito is truly from the
GRB 970417a.
I first try to estimate a likely value of the redshift and energetics of
this GRB from the observed energy fluence and detection rate (1 per 54
GRBs), under the assumption that this GRB is not a peculiar GRB.
I then discuss whether this extreme phenomenon
can be explained in a reasonable theoretical framework. 
I show that the proton-synchrotron model of GRBs (Totani 1998b, 1999)
gives a possible explanation for the Milagrito result.

Throughout this letter I use the isotropic energy for the
total energy of GRBs. My analysis does not depend on the unknown
collimation factor of GRBs, and the actual energy emitted by
the central engine can be much smaller than the isotropic energy
if GRBs are strongly collimated, jet-like explosions.

\section{Redshift and Energetics of GRB 970417a}
The basic assumption is that GRB 970417a
is the nearest GRB to us among the 54 GRBs observed by the Milagrito.
The unknown luminosity function of GRBs in the TeV range 
makes this assumption less reliable, but
this is reasonable because detectability of TeV
gamma-rays from cosmological GRBs very rapidly decreases with increasing
redshift, due to the well-known effect of intergalactic absorption
by the cosmic infrared background radiation 
(see, e.g., Salamon \& Stecker 1998;
Primack et al. 1999). Another argument supporting this assumption
will be given in \S 4.

We can estimate the fraction of GRBs within a given redshift $z$ of
all GRBs detectable by the BATSE satellite, $N(<z)$, if we know 
the sub-MeV luminosity
function of GRBs, GRB rate history as a function of $z$, and
the threshold flux of the BATSE. Let $L_p$ be the peak photon-number
luminosity (i.e., not energy)
in the restframe 50--300 keV range, and $P$ be the observed
peak photon flux in the BATSE range of 50--300 keV. These two are
related as:
\begin{equation}
P(L_p, z) = \frac{ (1 + z)^{2-\alpha}}{4 \pi d_L^2} L_p \ ,
\end{equation}
where $d_L$ is the standard luminosity distance and 
I have assumed a power-law spectrum of GRBs in the sub-MeV range
with a spectral index $\alpha$,
as $dL_p/d\varepsilon \propto \varepsilon^{- \alpha}$. 
The observed rate of GRBs with a given set of $(L_p, z)$ is
given as
\begin{equation}
\frac{d^2N}{dL_p dz} = \phi(L_P) \frac{R_{\rm GRB}(z)}{(1+z)}
\frac{dV}{dz} \ ,
\end{equation}
where $\phi(L_p)$ is the luminosity function of GRBs, $R_{\rm GRB}$
is the comoving rate density of GRBs as a function of redshift,
and $dV/dz$ is the comoving volume element of the universe.
The factor $(1+z)^{-1}$ comes from the cosmological time dilation effect.
I assume that the GRB rate evolution traces the star formation
history in the universe (Totani 1997). 
The form of the star formation rate evolution
is modeled based on the data compiled by Madau, Pozzetti, \&
Dickinson (1998).
Then $N(<z)$ is given as
\begin{equation}
N(<z) = \frac{ \int_0^z dz \int dL_p (d^2N/dL_p dz)
\Theta [ P(L_p, z) - P_{\rm th} ] }{
\int_0^\infty dz \int dL_p (d^2N/dL_p dz)
\Theta [ P(L_p, z) - P_{\rm th} ]
}\ ,
\end{equation}
where $\Theta$ is the step function [i.e., $\Theta(x) = 1$ and 0
for $x>0$ and $x<0$, respectively], and $P_{\rm th}$ is the detection
threshold of the BATSE ($P_{\rm th} \sim 0.3 \ \rm photons \ 
cm^{-2} sec^{-1}$,
Meegan et al. 1996). In this letter I assume a form
of the luminosity function of GRBs as logarithmically flat 
[$\phi(L_p) \propto L_p^{-1}$] within a range of $(L_{p, \min},
L_{p, \max}) = (10^{57}, 10^{59})$ [photons s$^{-1}$], 
based on the observed luminosity distribution
of GRBs with secure redshifts (see, e.g., Table 1 of
Lamb \& Reichart 2000).
The spectral index $\alpha$ is set to 1, which is a rough average of
GRB spectra (Mallozzi, Pendleton, \& Paciesas 1996).
I use a reasonable set of cosmological parameters of
$H_0$ = 70 km/s/Mpc and $(\Omega_0, \Omega_\Lambda) = (0.3, 0.7)$.

\begin{figure}
  \begin{center}
    \leavevmode\psfig{file=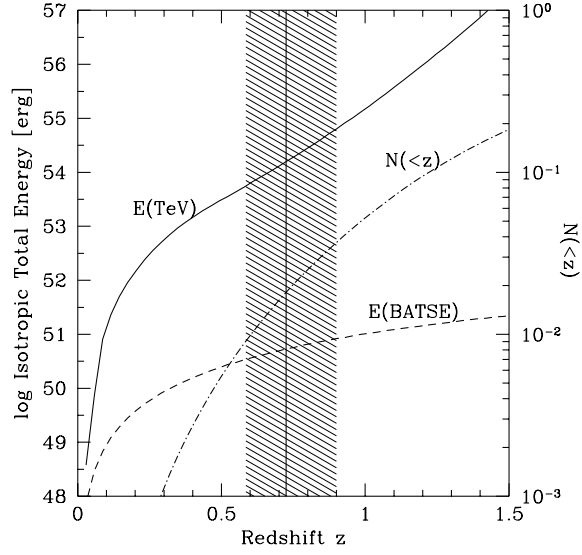,width=8cm}
  \end{center}
\caption{\footnotesize
The dot-dashed line shows the fraction of GRBs within 
a given redshift of all the BATSE GRBs, $N(<z)$, assuming logarithmically
flat luminosity function of GRBs and GRB rate evolution proportional to
the star formation rate in the universe (see text). See the
right-hand-side scale for $N(<z)$. The shaded region
shows a redshift range which is consistent with the detection rate
of GRBs in the TeV range by the Milagrito, 1 per 54 BATSE GRBs, within a 
factor of two. The vertical solid line in the center of the shaded
region indicates the redshift exactly consistent with the detection 
rate of 1/54. The solid and dashed 
lines are the estimates of total isotropic energy emitted from
GRB 970417a in the TeV and BATSE ranges, respectively. 
The intergalactic absorption
of TeV gamma-rays is taken into account for the total TeV energy.
}
\label{fig:E-z}
\end{figure}

In Fig. \ref{fig:E-z}, I show $N(<z)$ as a function of redshift
by the dot-dashed line. [See the right-hand-axis for the scale of $N(<z)$].
It becomes consistent with the detection rate of GRBs by
the Milagrito detector, 1/54, at redshift $\sim 0.7$.
(The vertical solid line shows the redshift corresponding to the 
detection rate of 1/54, and the shaded region shows 
a redshift range in which the detection rate is consistent 
within a factor of two.)
Also shown by the dashed line is the total isotropic energy emitted in the
BATSE range, $E_{\rm MeV}$, which is estimated by the BATSE fluence
of the GRB 970417a, $3.9 \times 10^{-7} \ \rm erg \ cm^{-2}$,
in all the four energy channels of the BATSE
($>20$ keV). [The latest BATSE catalog at MSFC is available at
http://cossc.gsfc.nasa.gov/cossc/BATSE.html].  
The isotropic energy in the sub-MeV band then becomes $\ltilde 10^{51}$
erg in the likely redshift range. The observed
total sub-MeV energy of the GRBs 
with secure redshifts is widely distributed in a range 
$\sim 10^{51}$--$10^{54}$ erg, and 
GRB 970417a belongs to a class of the least energetic GRBs 
in the sub-MeV band.

\begin{figure}
  \begin{center}
    \leavevmode\psfig{file=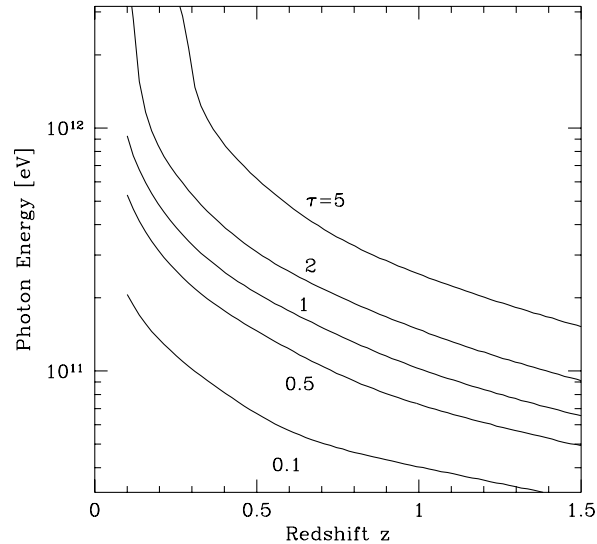,width=8cm}
  \end{center}
\caption{\footnotesize
Contour map of the intergalactic optical depth of
very high energy gamma-rays, as a function of observed
photon energy and source redshift. The value of optical 
depth, $\tau$, is indicated in the figure.
}
\label{fig:tau}
\end{figure}

In order to estimate the total energy emitted in the TeV range,
the absorption optical depth of
TeV gamma-rays due to the cosmic infrared background is necessary.
I have calculated this optical depth as a function of the source
redshift and observed photon energy, by using a 
model of luminosity density evolution 
of stellar lights in the universe (Totani, Yoshii, \& Sato 1997). 
The standard formulation for the calculation of optical depth
from the luminosity density evolution in the universe is given
in e.g., Salamon \& Stecker (1998). The dust-emission component 
is calculated assuming that the dust emission spectrum is the same
as that in the solar neighborhood and the fraction of stellar light
absorbed by dust is determined to reproduce the far infrared
background radiation measured by the COBE satellite
(Hauser et al. 1998). I have checked that this model of
optical depth is quantitatively consistent with earlier publications
within the model uncertainties (e.g., Salamon \& Stecker 1998;
Primack et al. 1999). Fig. \ref{fig:tau} shows the
observed photon energy ($\varepsilon$)
corresponding to several values of optical
depth [$\tau(z, \varepsilon)$] as a function of the source redshift.

The events observed by the Milagrito are considered to be
gamma-rays above 50 GeV (Atkins et al. 2000), and an estimate
of the TeV fluence of the GRB 970417a
is given as a function of the upper cut-off
energy and spectral index (see Fig. 4 of Atkins et al. 2000).
Here I assume the spectral index of 1.5, which is the standard
photon index of synchrotron radiation with particle index of 2.
If the TeV spectrum is harder or softer than this, the estimate of 
the total energy in TeV range becomes smaller or larger, respectively
(see Fig. 4 of Atkins et al. 2000).
I use the photon energy corresponding to the intergalactic 
optical depth $\tau$=1,
denoted as $\varepsilon_{\tau = 1}(z)$, as the upper cut-off
energy, unless this energy is smaller than 50 GeV. Then we can estimate
the total isotropic energy emitted in the TeV range as
\begin{equation}
E_{\rm TeV} = \frac{4 \pi d_L^2}{(1+z)} F_{\rm TeV}(\varepsilon_{\rm
cut}) \exp[ \tau(z, \varepsilon_{\rm cut}) ] \ ,
\end{equation}
where the upper cut-off energy is $\varepsilon_{\rm cut} = \max(
{\rm 50 \ GeV},  \varepsilon_{\tau=1})$, and $F_{\rm TeV}$ is
the observed TeV fluence which is a function of the upper cut-off 
energy. The result is shown by the solid line
in Fig. \ref{fig:E-z} as a function of redshift. 

From these results, a possible interpretation is that GRB 970417a
was located at $z \sim$ 0.7 emitting isotropic energies of
$\gtilde 10^{54}$ and
$\ltilde 10^{51}$ erg in the TeV and sub-MeV range, respectively.
It is interesting to note that the energy emitted in the TeV range
is similar to that in the sub-MeV range of the most energetic GRB
observed to date: GRB 990123, whose total isotropic energy
was estimated as $\sim 3 \times 10^{54}$ erg (Kulkarni et al. 1999). 
Therefore the extremely large isotropic energy 
in the TeV range for GRB 970417a is not too large for the total
energy budget of GRBs. In fact, we do not know
why most of the energy of GRBs is emitted in the sub-MeV
range, and there is no robust theoretical argument which excludes
a possibility that most of the total GRB energy is emitted in
other photon energy bands. In fact, such an extreme phenomenon has been
predicted by the proton-synchrotron model of GRBs (Totani 1998b; 1999). 
In the rest of this letter I discuss
whether this model can explain the Milagrito result.

\section{Interpretation by the Proton-Synchrotron Model}
Full description of this model has already been given in
Totani (1998b; 1999), and here I summarize the qualitative feature of
the model. 

Currently the most popular explanation for the GRB phenomenon
is dissipation of the kinetic energy of ultra-relativistic bulk motion
with a Lorentz factor of $\Gamma \gtilde 10^{2-3}$, in internal shocks
which are generated by relative velocity differences of relativistic
shells ejected from the central engine.
All the total energy ejected as relativistic bulk motion cannot be
dissipated in internal shocks, and hence the total energy truly emitted as
kinetic motion ($E_{\rm iso}$)
should be larger than the observed total energy of gamma-rays,
$E_{\rm \gamma,iso}$, at
least by a factor of several. Therefore, 
the most energetic class of GRBs, such as GRB 990123,
must emit quite a large amount of energy, $E_{\rm iso} \gtilde
10^{55}$ erg. If the efficiency of the internal shock is not so high,
we may have to consider an isotropic energy reaching $\sim 10^{56}$ erg.
Therefore, if GRBs are produced by stellar death events, GRBs must be
strongly collimated at least by a factor of ($4 \pi / \Delta \Omega) \sim$
100 to reduce the actual energy emitted from the central engine.

Since the origin of the GRB energy is relativistic bulk motion,
protons should carry a much larger amount of energy than electrons
by a factor of $m_p/m_e \sim 2,000$
in the initial stage of the internal shock generation.
It is very uncertain what fraction of the proton energy is converted into
electrons, but the simplest Coulomb interaction cannot 
transfer the proton energy into electrons within the typical time scale of GRBs
(Totani 1998a, 1999). The soft gamma-rays are generally considered
to be generated by electrons, because of the short time variability of GRBs.
Therefore it is not unreasonable
that, in some GRBs, only $(m_e/m_p) \sim 10^{-3}$ of the total kinetic energy
is carried by electrons and then emitted as soft gamma-rays. 
If the hidden energy carried by protons is directly emitted in
the TeV range, then much more energy can be radiated in the TeV range
than in the sub-MeV range by a factor of almost one thousand.

GRBs are known as a possible site for the acceleration of
protons up to $10^{20}$ eV, which are observed on the Earth
as ultra-high energy cosmic rays (Waxman 1995; Vietri 1995;
Milgrom \& Usov 1995).
I have already shown (Totani 1998b) that, 
when $E_{\rm iso} \gtilde 10^{55}$ erg and the magnetic field
is as strong as the energy density of the shocked region, synchrotron
radiation of protons accelerated to $10^{20}$ eV can be an efficient
emission process because the cooling time of such protons is comparable with
the typical GRB duration ($\sim$ 10 sec) in the observer's frame. The energy of
these synchrotron photons for an observer is about 1--10 TeV, and 
strong TeV emission from GRBs is possible. I suggest that
the TeV gamma-rays possibly detected by the Milagrito were
produced by this mechanism.

On the other hand, it may also be possible that a physical process works as
an energy conveyor from the hidden energy reservoir (i.e., protons) into
electrons (or positrons). If the energy transfer is almost complete in a GRB, 
a significant fraction of $E_{\rm iso}$ can be radiated as gamma-rays
in the sub-MeV range.
I have pointed out (Totani 1999) that 
$e^\pm$-pair creation by TeV photons of proton-synchrotron might
work as the new energy channel for the energy transfer from protons
into electrons and positrons, giving an explanation for
the energetic sub-MeV GRB phenomenon such as GRB 990123. 
Proton-synchrotron photons interact with low energy
electron-synchrotron photons and create $e^\pm$ pairs. 
It can also be shown that the photon energy range of the synchrotron
radiation of the created pairs becomes about MeV, i.e., consistent
with the BATSE range.

Then what is the crucial parameter which determines whether a
GRB is bright in TeV or MeV?  The GRB luminosity in the sub-MeV range
is determined by the efficiency of energy transfer
from protons into $e^\pm$ pairs, i.e., the opacity of pair-production 
reaction for the proton-synchrotron TeV photons.  
Based on the typical fireball parameters of GRBs, this pair-production opacity 
is typically of order unity, and strongly depends on the bulk
Lorentz factor of GRBs by the special relativistic effect
as $\tau \propto \Gamma^{-5}$ in a simple internal shock model
(Totani 1999; see also Baring \& Harding 1997; B\"ottecher \& Dermer 1998).
Because of this strong dependence of the pair-creation optical depth
on $\Gamma$, a modest dispersion in $\Gamma$ by a factor of 3--4
from one GRB to another results in
drastic change in the sub-MeV energetics of GRBs by a factor of up to
$\sim 10^3$ (Totani 1999).
A large value of $\Gamma$ results in negligible optical depth to
pair-creation and hence a GRB strong in TeV such as GRB 970417a, while a small 
$\Gamma$ in the inverse case of a GRB strong in MeV such as GRB 990123.

This mechanism gives a natural explanation for the wide dispersion
in the observed total GRB energy in the sub-MeV range, with almost
no correlation with the afterglow luminosity (Totani 1999). This model assumes 
a relatively
uniform distribution of the total kinetic energy emitted from the
central engine, and sub-MeV luminosity of GRBs is not correlated with
the kinetic energy injected into interstellar/circumstellar medium.
It may be similar to a see-saw between sub-MeV and TeV energies, 
in which the total kinetic
energy of GRBs is roughly the same for all GRBs and difference of GRB
energetics is whether dominant emission is in TeV or MeV bands.

To summarize, GRB 970417a observed by the Milagrito can be 
understood as a GRB with the isotropic kinetic energy $E_{\rm iso} \gtilde
10^{55}$ erg ejected from the central engine,
a significant fraction of which is radiated in the TeV range
by the synchrotron radiation of ultra-high-energy protons,
with almost no energy transfer from protons into electrons
due to a relatively large value of $\Gamma$.

\section{Discussion}
Here I discuss some observational signatures expected in future experiments
from the interpretation presented in this letter. 

The proton-synchrotron
model predicts that the ratio of TeV/MeV luminosities drastically changes
from burst to burst by the difference of energy transfer efficiency
from protons into electrons and positrons. When the optical depth
to the pair-creation reaction is much larger than unity,
it is possible that the TeV energy fluence is much weaker than
the sub-MeV fluence in contrast to the GRB 970417a, 
even after the intergalactic absorption of 
TeV gamma-rays is corrected. However, a generic prediction of this model
is that the total of isotropic TeV and sub-MeV
energies is about $\sim 10^{54-55}$ erg for all GRBs with much
smaller scatter than that of the sub-MeV or TeV isotropic energies.

The assumption that the GRB 970417a is the closest burst to us
may be wrong if TeV luminosity of GRBs drastically changes from burst to
burst, as expected in the proton-synchrotron model. However, I emphasize that 
this assumption is, in this case, conservative from a theoretical
point of view. The argument that the redshift of the closest GRB
in the 54 BATSE GRBs is about $z \sim 0.7$ is based only on the sub-MeV
luminosity function of GRBs and GRB rate evolution.
Then, if the GRB 970417a is not the closest but
intrinsically even brighter GRB in TeV,
its distance must be larger than $z \sim 0.7$. Therefore
this estimate can be considered as a lower limit of the redshift.
If the redshift is significantly larger than 0.7,
the TeV isotropic energy of this burst would be much larger than $\sim
10^{54}$ erg, which would be quite difficult to explain by
the stellar death models even if we invoke a strongly collimated
jet-like explosion.  Hence I consider the estimate
of redshift and energetics presented here is reasonable.

The Milagro detector, which has significantly
increased the sensitivity to GRBs between 0.1 and 10 TeV, is now operating
to search for GRBs (Atkins et al. 2000). 
If the signal observed by
the Milagrito is truly from GRB 970417a, the Milagro detector would
detect an event similar to GRB 970417a with better signal-to-noise ratio.
It is also important to detect GRBs more distant than GRB 970417a
to increase the detection rate, because $N(<z)$ rapidly increases
with redshift. It is unfortunate that the intergalactic
optical depth of TeV gamma-rays also rapidly increases, and hence
improvement of detector sensitivity does not significantly extend
the distance of marginally detectable GRBs. 
The cut-off photon energy by the intergalactic absorption
falls below $\sim$ 50 GeV at $z \sim 1.2$, where $N(>z) \sim 0.1$.
Therefore it seems unlikely that the detection rate
in the TeV range is increased to more than
1 per 10 BATSE-detected GRBs, even if the detector sensitivity 
is significantly improved.

Another improvement of the Milagro over Milagrito is the improved
of spectral information. If we can measure the spectral cut-off
by the intergalactic absorption for GRBs with known redshifts,
it would provide information on the flux of the cosmic
infrared background radiation as well as 
its evolution to $z \gtilde 0.5$, which 
would be valuable for the study of galaxy formation and evolution.

\newpage

\end{document}